\documentclass{iopart}
\usepackage{graphicx}

\begin{document}

\newcommand{\remark}[1]{{\large\bf #1}}
   
\title{Ground states of $2d$ $\pm J$ Ising spin glasses via stationary Fokker-Planck sampling}
\author{O Melchert and A K Hartmann}
\address{Institut f\"ur Physik, 
Universit\"at Oldenburg, 
Carl-von-Ossietzky Strasse, 
26111 Oldenburg, Germany}
\ead{melchert@theorie.physik.uni-oldenburg.de and alexander.hartmann@uni-oldenburg.de}

\begin{abstract}
We investigate the performance of the recently 
proposed stationary Fokker-Planck 
sampling method considering a combinatorial optimization problem 
from statistical physics. 
The algorithmic procedure relies upon the 
numerical solution of a linear 
second order differential equation that depends on a diffusion-like 
parameter $D$.
We apply it to the problem of finding ground states
of $2d$ Ising spin glasses for the $\pm {\rm J}-$Model. We consider 
square lattices with side length up to $L=24$ with two different 
types of boundary conditions
and compare the results to those obtained by exact methods.
 A particular value of $D$ is found that yields an optimal performance
of the algorithm. 
We compare this optimal value of $D$ to
a percolation transition, which occurs when studying the connected
clusters of spins flipped by the algorithm. 
Nevertheless, even for moderate lattice sizes,
the algorithm has more and more problems to find the exact ground states. 
This means that the approach, at least in its standard form,
seems to be inferior to other approaches like parallel tempering.
\end{abstract} 

\pacs{75.40.Mg, 02.60.Pn, 02.50.-r}

\section{Introduction \label{sect:introduction}}

During the last decades, a vast number of heuristic  methods where
developed that aim to solve optimization  problems by employing ideas
from physics and related  disciplines
\cite{opt-phys2001,opt-phys2004}.  Among those, stochastic search
strategies like e.g.\  simulated annealing
\cite{kirkpatrick1983,schneider2006},  parallel tempering Monte Carlo
\cite{geyer1991,hukushima1996}, extremal  optimization
\cite{boettcher2001,middleton2004} and genetic algorithms
\cite{pal1996,hartmann1999}  provide valuable tools to locate points
in the configuration space that correspond to a near-optimal or  even
an optimal value of the underlying cost function.  Here we investigate
a recently proposed heuristic for stochastic optimization, called
stationary Fokker-Planck (SFP) sampling
\cite{berrones2008a,berrones2008b}.  The basic idea is to
perform a Langevin dynamics for the variables of a given system in the
potential given by the cost function. Through iteratively decreasing
the stochastic noise, the variables shall be driven into a global
minimum of the cost function.  Langevin dynamics can also be cast in
terms of a  Fokker-Planck equation as evolution equation for the
probability density of the variables of the cost function.
Related to this, SFP sampling aims to estimate the asymptotic
probability  density of a stochastic search process by estimating the
marginal  densities of the individual variables that enter the cost
function.  Moreover, the  influence of the cost function on the search
process therein depends  on a diffusion-like parameter $D$.
Upon its introduction in \cite{berrones2007}, the approximation of a
stationary probability density by the SFP algorithm was illustrated
for the 2 parameter Michalewicz function, an unconstrained test
function for global optimization.
In \cite{pena2007} the authors presented an implementation of the SFP
algorithm and showed its applicability to the 5 parameter Levy No. 5
function, again a test function for global optimization and the XOR
problem, a fundamental problem e.g. relevant to the subject of machine
learning. In both latter  cases the SFP algorithm was used to
construct probability densities, consistent  with the asymptotic
statistical properties of the search process. The point in  the search
space that was found to have the maximum probability was then used as
an initial point for the search of a global optimal point using the
deterministic powell algorithm \cite{numrec}.

We now give a brief overview over the past applications of the
SFP algorithm.  In \cite{berrones2008a} the SFP algorithm was
detailed further and tested for  various unconstrained optimization
problems. Among those were the 2 parameter  Schwefel function, a
separable problem for which the algorithm converges  after only one
step of iteration and the 20 parameter Rosenbrock function.
Originally designed for problems defined on continuous  unconstrained
configuration spaces, the SFP algorithm can also be applied  to
discrete and constrained problems by the introduction of  proper
penalty functions in addition to the cost function.  The introduction
of additional functions intended to model constraints on the
individual variables was illustrated for the knapsack problem, a
NP-hard combinatorial  optimization problem. Therefore, instances of
the knapsack problem with up to 30  variables where
considered. Besides that, the construction of heuristics based  on the
SFP algorithm in combination with a downhill simplex routine
\cite{numrec} was discussed.
Recently \cite{berrones2008b} the asymptotic convergence properties of
the SFP algorithm were studied by means of numerical experiments
considering the 2 parameter Michaelewicz function and the XOR
problem. Moreover, its applicability to problems that arise in the
field of statistical inference was outlined by showing how the SFP
algorithm can efficiently be used to perform maximum likelihood and
Bayesian training of neural networks.

For all these cited applications, system with rather few
degrees of freedom were studied. To get an impression, whether an
optimization algorithm is really competitive, larger problems have to
be treated.  Here, we investigate the performance of the SFP
algorithm by applying it to a  combinatorial optimization
problem from statistical physics.  More precise, we perform ground
state (GS) calculations of two-dimensional $\pm J$ Ising spin glasses
(ISGs) with nearest-neighbor  interactions. The $\pm J$ ISG is a
disordered model system in which the sign of the interactions is drawn
randomly but the magnitude  of the interactions is fixed to a value
$J$.  We consider square lattices of side length $L\leq 24$,  i.e. up
to $N=576$ spin variables.  The fact that there exist exact algorithms
that yield GS  properties within a computing time that is bounded by a
polynomial  in the number of spins, turns the $2d$ ISG into an
expedient testbed  to evaluate the performance of heuristics like the
SFP algorithm.
Albeit the task of the SFP algorithm is to identify regions in the search 
space that contain optimal values of the cost function with high probability,
our intention here was to check as to which extend the SFP algorithm 
can be used as an heuristic in order to identify ground states of 2d $\pm J$ spin
glasses.\\
The paper is organized as follows. In section \ref{sect:model} we
describe $\pm J$ ISGs in more detail, in section  \ref{sect:algorithm}
we explain the heuristic algorithm that we have employed  in order to
find GS spin configurations and section \ref{sect:results}  contains
the results on the algorithm performance.  Section
\ref{sect:conclusions} concludes with a summary.

\begin{figure}[t!]
\centerline{ \includegraphics[width=0.7\linewidth]{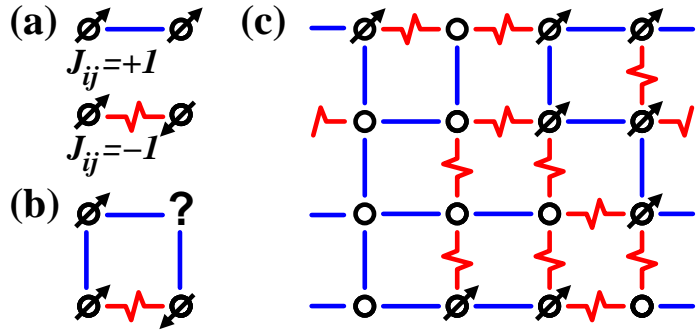}}
\caption{Features of an ISG: (a) ferromagnetic
($J_{ij}=+1$, straight line) and antiferromagnetic  ($J_{ij}=-1$,
jagged line) bonds. Ferromagnetic bonds favor parallel
alignment of the connected spins, while an antiferromagnetic bond
leads preferentially to an antiparallel alignment.  (b) Competing
interactions give rise to frustration: no matter how the spin at the
upper right corner is oriented, one  of the incident bonds is not
satisfied.  (c) One GS for an instance of a $\pm J$ ISG with $L\!=\!4$
and  periodic boundary conditions in the horizontal direction. For
clarity only those spins are shown that point upwards.
\label{fig:fig1abc}}
\end{figure}  

\section{Model\label{sect:model}}
%
%
Within the scope of this paper, we perform GS calculations of
two-dimensional Ising spin glasses with nearest-neighbor
interactions. The model consists of $N=L \times L$ spins
$\sigma=(\sigma_1,\ldots,\sigma_N)$ with $\sigma_i=\pm 1$ located on
the sites of a square lattice.  The energy of a given spin
configuration is measured by the Edwards-Anderson Hamiltonian
\begin{eqnarray}
H(\sigma) = -\sum_{\langle i,j \rangle} J_{ij} ~\sigma_i \sigma_j,
\label{eq:EA_hamiltonian}
\end{eqnarray}
where the sum runs over all pairs of adjacent spins.  Therein, the
bonds $J_{ij}$ are quenched random variables drawn from a bimodal
disorder distribution that allows for values  $J_{ij}=\pm 1$ with
equal probability ($\pm J$--model).  The bonds can take either sign
and thus lead to competing interactions  among the spins giving rise
to frustration, see figure \ref{fig:fig1abc}.  A plaquette, i.e. an
elementary square on the lattice, is said to be frustrated if it is
bordered by an odd number of negative bonds.  Frustration, in effect,
rules out a GS in which all the bonds are  satisfied.  For the $\pm
J$--model considered here, the GS is highly degenerate: there are
numerous spin configurations that all have minimal energy. Moreover,
the average number of GSs increases  exponentially with $N$
\cite{saul1993,landry2002}.  As a result, the problem of finding a GS
spin configuration for a  given realization of the bond disorder is
quite involved.  In the remainder of this section, we very briefly
describe the algorithms that we have used to generate the exact GS
data to which we compare the results of the SFP heuristic.

%
%
(i) free-periodic (FP) boundary conditions (BCs): for the $2d$ ISG,
where there are periodic boundary  conditions in at most one
direction, exact GS spin configurations can be found in polynomial
time.  This is possible through a mapping to an appropriate minimum
weight perfect matching problem
\cite{opt-phys2001,bieche1980,SG2dReview2007}. 
Here we state only the general idea of this method. 
For this mapping, the system needs to be represented by its frustrated
plaquettes and paths connecting those pairwise, i.e.\
{\em matching} them. In doing so,
individual path segments are confined to run perpendicular across
bonds on the spin lattice.  Those bonds that are crossed by path
segments are not satisfied in  the corresponding spin configuration.
The {\em weight} of the matching is just the sum of all
unsatisfied bonds.
Hence, finding a minimum weighted perfect matching on the graph of frustrated
plaquettes then corresponds to finding a spin configuration  on the
original spin lattice for which a minimal number of bonds are broken
(in case of the $\pm J$--model), hence a GS. Using this
approach, since the minimum-weight perfect 
matching problem is polynomially solvable, 
rather large systems, easily with $L=500$ \cite{eScaling2004}, can be treated
on single processor systems.

(ii) periodic-periodic (PP) BCs: for the $2d$ ISG with fully periodic
boundary conditions, the  density of states can be calculated in
polynomial time  \cite{saul1993,saul1994}.  The respective algorithm
relies on a combinatorial expansion of  the partition function,
originally introduced for the pure $2d$  Ising model
\cite{kac1952}. Therein, the algorithmic procedure is based on a
high-temperature expansion of the underlying partition  function,
where it is possible to relate the coefficients of  individual
expansion terms to closed graphs on the square lattice.  Note that
this algorithm does not yield GS spin configurations,  instead it
gives rise to the GS energy. Due to the different
approach, somehow smaller systems $L\approx 64$ are easily feasible.

\section{Algorithm\label{sect:algorithm}}
Here, we briefly describe the stationary Fokker-Planck sampling
algorithm as  introduced in \cite{berrones2007} and further detailed
in \cite{pena2007,berrones2008a,berrones2008b},  where also more
details about the general framework can be found. SFP sampling is
based on the interrelation between the Langevin and Fokker-Planck
equations that allow for the stochastic description  of a given
system.  As noted in \cite{berrones2008a} a Langevin equation, as
stochastic differential  equation, contains the building blocks for a
stochastic search strategy.  In this regard, a stochastic dynamics for
a set of variables $x=(x_1,\ldots,x_N)$ under  the influence of a cost
function $V(x)$ can be written as
\begin{eqnarray}
\dot{x}_n=-\frac{\partial V(x)}{\partial x_n} + \epsilon(t),
\end{eqnarray}
where $\epsilon(t)$ is a Gaussian white noise with mean $\langle
\epsilon(t) \rangle=0$ and a correlation function $\langle
\epsilon(t)\epsilon(t') \rangle=D \delta(t-t')$, where $D$ signifies
the diffusion constant.  By the above equation  the variables $x_n$
interact through the forces induced by the cost function $V(x)$ under
the influence of a rapidly fluctuating noise $\epsilon(t)$.  Further,
the probability density of a system as modeled by the equation above
is governed by a Fokker-Planck equation of the form
\begin{eqnarray}
\dot{p}(x) = \sum_{n=1}^N \frac{\partial}{\partial x_n} \Big[
\frac{\partial V}{\partial x_n} p(x) \Big] + \sum_{n=1}^N \sum_{m=1}^N
D_{nm} \frac{\partial^2 p(x) }{\partial x_n \partial x_m },
\label{eq:FP}
\end{eqnarray}
a linear differential equation, where $D_{nm}=\delta_{nm} D$.  The
absence of infinite cost values and a bounded search space for the
optimization of $V(x)$, realized by constraints of the form
$L^{-}_n\leq x_n \leq L^{+}_n$ further ensures that a stationary
solution for the density $p(x)$ exists.  One introduces the conditional
densities
\begin{eqnarray}
p(x_n | \{x_{j\neq n}\})=\frac{p(x)}{\int p(\{x_{j\neq n}\}|x_n)p(x_n)
dx_n},
\end{eqnarray}
i.e. the evolution of variable $x_n$ given the positions of all the
other variables where $p(x_n)=\int p(x_n|\{x_{j\neq n}\})p(\{x_{j\neq
n}\})d\{x_{j\neq n}\}$ denotes the respective marginal density. The
one-dimensional projection of equation (\ref{eq:FP})
\begin{eqnarray}  
D \frac{\partial p(x_n |\{x_{j\neq n}=x_j^*\}) }{\partial x_n} + p(x_n
| \{x_{j\neq n}=x_j^*\}) \frac{\partial V(x)}{\partial x_n}=0,
\label{eq:FP_1dim_projection}
\end{eqnarray}
describing the evolution of variable $x_n$ while keeping all the other
variables  at fixed positions $x_{j\neq n} = x_j^*$, can be used to
draw points from  the conditional density and therewith sample the
corresponding marginal density $p(x_n)$.  This is possible by casting
equation (\ref{eq:FP_1dim_projection}) into a linear second order
differential equation for the cumulative distribution $y$ that is
connected to the marginal density via $y(x_n|\{x_{j\neq
n}=x_j^*\})=\int_{-\infty}^{x'_n} p(x_n | \{x_{j\neq n}=x_j^*\})
dx'_n$:
\begin{eqnarray}
\frac{d^2 y}{dx_n^2} + D^{-1} \frac{\partial V}{\partial x_n}
\frac{dy}{dx_n} =0,
\label{eq:cum_1dim}
\end{eqnarray}
together with the boundary conditions $y(L^-_n)=0$ and  $y(L^+_n)=1$,
$y$ is an uniformly distributed random variable in the interval
$[0,1]$.  Therefore an inversion method can be used to draw random
deviates from the conditional density.  Further, the iterative
execution of the following steps yield an algorithm for the
approximation of $y$:\\ (1) Fix the variables $x_{j\neq n}=x_j^*$ and
approximate $y(x_n|\{x_{j\neq n}\})$ by  use of equation
(\ref{eq:cum_1dim}).\\ (2) Construct a lookup table from
$y(x_n|\{x_{j\neq n}\})$ so as to generate a deviate $x_n^*$ drawn
from the stationary distribution $p(x_n|\{x_{j\neq n}=x_j^*\})$.\\ (3)
Update $x_n=x_n^*$ and repeat the procedure for a new variable
$x_{j\neq n}$.\\ Finally, the marginal $y(x_n)$ can be obtained as the
expected value of the conditional $y(x_n|\{x_{j\neq n}\})$ over the
set $\{x_{j\neq n}\}$ and  the marginal densities $p(x_n)$ for the
individual variables $x_n$ follow by differentiation.

The implementation of the spin model introduced in section
\ref{sect:model} is carried  out as follows. For a numerical treatment
of equation (\ref{eq:cum_1dim}) we set up its  finite difference
analogue based on a one-dimensional mesh with an over all number of  $m$
mesh points and a uniform mesh width $\Delta$. We therein employed a
second order  centered approximation for the derivatives as well as the
cost function in terms of the mesh width. The cost function is further
identified with the energy function  of the spin system, see equation
(\ref{eq:EA_hamiltonian}), and the set of variables $x$  is consequently
identified with the spin degrees of freedom that are bounded by
$-1\leq \sigma_n \leq 1$ and are allowed to take one of the $m$ values
$-1, -1+\Delta, \ldots, 1-\Delta, 1$.  However, according to the model
introduced in section \ref{sect:model} valid configurations consist of
spins with integer values $\pm 1$ only. To account for these
constraints on the spins, the energy function gets an additional term
that favors those configurations, where the  constraints on the spins
are met. E.g.\ the local energy of an  individual spin can be changed
to
\begin{eqnarray}
V(\sigma_i|\{\sigma_{j\neq i}\})= -\sum_{j\in N(i)} J_{ij} \sigma_j
\sigma_i ~+~\lambda(1-\sigma_i \sigma_i), \label{eq:localErg}
\end{eqnarray}
wherein $N(i)$ comprises all spins adjacent to spin $i$.  In this
case, any deviation from the allowed spin values is  payed off with an
additional cost, while the penalty term equates to zero if the
constraints are met.  
%
Initially we performed numerical experiments with different types of penalty
terms. Finally we decided to use the present one because it is the
most basic one can think of, e.g. it involves only one parameter. 
One thing to note about the energy function of the Ising model is that,
regardless of the system size, the local energy of an individual spin can only
take values in between $\pm4J$ (for the 2d model). This means, once an 
optimal parameter for the penalty term is found it should not change with the
system size.
In our initial experiments we found that the effect of the penalty term is
best, if the magnitute of the penalty term is approximately of the order of
the local energy. It does not matter if its somewhat smaller/larger but the 
penalty term should not dominate the local energy of an individual spin. 
We fine tuned the penalty term for a system of a given size but we did not 
check whether the value changes for smaller/larger systems because of the 
reasons above.
During the execution of
the algorithm, we measure configurational properties  like energy and
magnetization $M=\sum_i \sigma_i /N$ that correspond to the
momentarily approximated density. Despite the penalty term in equation
(\ref{eq:localErg}),  the momentarily density can lead to spin
configurations where the $\sigma_i$ are not strictly $\pm 1$ but
deviates slightly. We
correct for those deviations by rounding to the closest number 
$\pm 1$, before we evaluate configurational properties.
In the following we present our results on the performance of the SFP
algorithm applied to the problem of finding ground states of $2d$ $\pm
J$ ISGs.


\section{Results \label{sect:results}}

So as to quantify the performance of the SFP heuristic we compare the
resulting GS spin configurations for $2d$ $\pm J$ ISGs to those
obtained by the exact methods outlined in section \ref{sect:model}.
Specifically,
 we generated $100$ test instances for several system sizes
in  the range $L=4,\ldots,24$ for both FP and PP BCs.  One such test
instance consists of a set of bonds together with the  corresponding
GS energy. On input the SFP algorithm takes the set of bonds, the GS
energy, a value $D$ for the diffusion constant, a seed that  controls
steps (2) and (3) listed in section \ref{sect:algorithm} and  an over
all number of sweeps $t$ after which the respective run of the
algorithm should terminate.  One sweep therein consists of $N$
iterations  of the three steps introduced in section
\ref{sect:algorithm}.
%
%
\begin{figure}[t]
\centerline{ \includegraphics[width=0.7\linewidth]{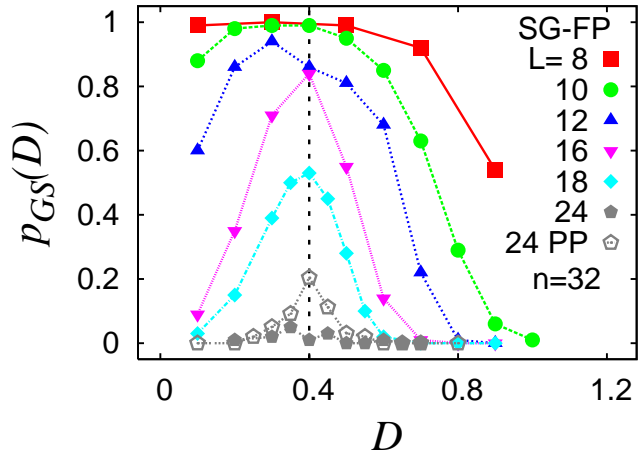}}
\caption{
Results for FP-BCs with a maximal number of
$t=10^4$ sweeps. The figure shows the fraction $p_{GS}$ of correctly identified GSs as
function of  the diffusion constant $D$ for different system sizes
$L$.
\label{fig:fig2}}
\end{figure}  
%
The spin
configuration from which a run of the SFP algorithm is started is
initialized at random. The algorithm constructs  an approximate density from
which the spin configurations are accessible. If the accordant
configurational energy turns out to be be equal to the GS energy the
run is completed and recorded as successful. If the computational
budget is exceeded and the  GS energy is 
not reached by any encountered spin
configuration then the run is  recorded as not successful. 
Finally, we  simulated a number of $n$ copies
of each test instance with different  initial spin configurations and
different seeds for the SFP algorithm.
For completeness, we mention that within the simulations we fixed the
value of $\lambda$ that enters the penalty function in equation
(\ref{eq:localErg}) to $\lambda=1/4$.
%
%
The performance of the algorithm also dependes on the number of mesh points 
that are used in solving equation (\ref{eq:cum_1dim}).
So as to find a proper number of mesh points $m$ there are two things to 
consider: on the one hand, the computational effort strongly depends on 
the number of mesh points, i.e. the larger the value of $m$ the more costly 
it is to solve the differential equation for the cumulative distribution; 
on the other hand, the number of mesh-points should nevertheless be large 
enough to ensure that the corresponding mesh-width resolves the cost function.
For the Ising spin system considered here, the only relevant part of the cost 
function is the local energy for the pivoting spin. We experienced that the 
local energy is rather smooth, allowing us to limit the number of mesh points 
to a value $m=40$. We fixed this value after some initial numerical experiments 
using different values in the intervall $m=10,...,120$. In particular at $D=0.4$ 
,i.e. the point where the algorithm perfomance is at its best, the results do 
not change for values $m>20$.
\subsection{Dependence on the diffusion constant}
We performed GS calculations using the SFP algorithm for different
values for the diffusion constant $D=0.1,\ldots,1.3$ and for different
system sizes $L$. For each test instance we simulated $n=32$ copies.
Consequently a test instance was completed successfully if at least
one of the $n$ runs correctly identified a GS.
%
\begin{figure}[t]
\centerline{ \includegraphics[width=0.7\linewidth]{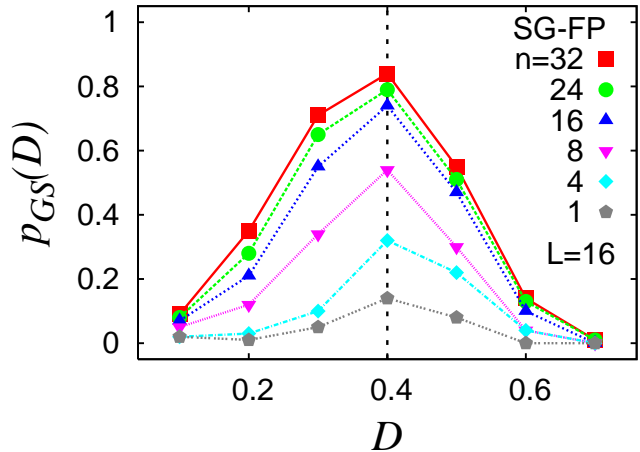}}
\caption{
Results for FP-BCs with a maximal number of
$t=10^4$ sweeps. The figure shows the effect of the number of 
simulated copies $n$ on the fraction
of correctly identified GSs. 
\label{fig:fig3}}
\end{figure}  
%
First of all we were interested in the fraction $p_{GS}$ of correctly
identified  GSs as a function of the diffusion constant. Results for
the FP-BC case  with $n=32$ and a number of $t=10^4$ sweeps are shown
in figure \ref{fig:fig2}.  For small system sizes ($L \leq 10$) the
algorithm correctly identifies GSs for a large  fraction
($p_{GS}=0.8$--$1.0$) of test instances over a comparatively broad
region of $D$ values.  For larger system sizes it performs best close
to $D=0.4$, as evident from figure \ref{fig:fig2}.
From equation (\ref{eq:cum_1dim}) it can be seen that in the limit of
large $D$  the influence of the cost function on the search process is
strongly suppressed.  Hence, in the limit $D\rightarrow \infty$ the
search process gets purely random.  On the other hand, at low values
of $D$ the cost function dominates the search process and the system
rapidly evolves towards a local minimum in the  configurational energy
and gets stuck there very likely.  At intermediate values of the
diffusion constant the search process is properly  guided by the cost
function and the algorithm has the ability to surmount energy barriers
in order to proceed towards a more feasible spin configuration.  To
support this intuition figure \ref{fig:fig5ab} illustrates three runs
of the  SFP algorithm for $D=0.1$, $0.4$ and $1.0$.  However, for a
fixed computational budget it can be seen that the GS-yield of  the
algorithm rapidly decreases with increasing system size. As
consequence we did not consider systems of size $L>24$.
This means, one of the main results of this paper, that the SFP
is not able to find exact GSs for even moderate sizes
of the system.
%
%
\begin{figure}[t]
\centerline{ \includegraphics[width=0.7\linewidth]{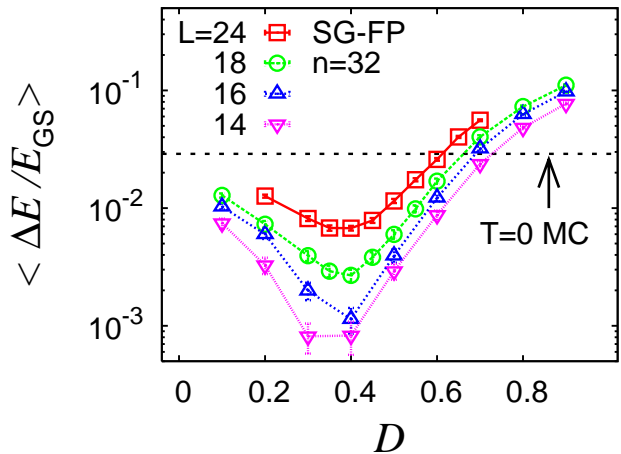}}
\caption{
Results for FP-BCs with a maximal number of
$t=10^4$ sweeps. The figure shows the relative difference of the best found
energy to the GS energy as function of $D$. The horizontal dashed line
indicates the respective value obtained using a $T=0$ MC simulation.
\label{fig:fig4}}
\end{figure}  
%

Figure \ref{fig:fig3} shows how the value of $p_{GS}$ is
influenced  by the number $n$ of simulated copies of an test instance.
Albeit the advantage decreases with increasing $n$, the data suggests
that it is beneficial to perform more than just a single run on a
given  test instance.  To understand this issue, we investigated the
distribution of the time until termination of the SFP algorithm for an
individual test instance more  closely. More precise, we performed
$10^6$ runs for one test instance of size $L=8$, again for different
values of $D$.  We found a heavy-tailed distribution, peaked at a
small value of $t$ ($t\approx 30$ for $D=0.4$), where for  larger
values of $D$ the peak is less pronounced. 

In cases where the algorithm fails to identify a GS correctly, it
nevertheless reaches a spin configuration with an energy $E = E_{\rm
GS} + \Delta E$ close to optimality as can  be seen from the averaged
relative difference $\langle \Delta E / E_{\rm GS} \rangle$  to the GS
energy, illustrated in figure \ref{fig:fig4}.  Therein, the best
performance of the algorithm is again obtained for a value close to
$D=0.4$.
The horizontal line indicates the respective value obtained using a
zero temperature Monte Carlo ($T=0$ MC) simulation: While
starting with a random initial spin configuration proposed spin flipps
are accepted only if this reduces the  configurational energy. This is
the most naive algorithm one can devise and it almost surely evolves
towards a local minimum. As evident from figure \ref{fig:fig4},
the SFP algorithm at $D\approx0.4$ outperforms the $T=0$ MC simulation
but at values $D> 0.6$ the latter one appears to perform better. This
is due to the fact that at higher values of $D$ the search process is
no longer guided  by the cost function in a proper manner.
Also one can see, that for $D\to 0$, the results of the SFP
algorithm converges towards the $T=0$ MC result, as expected.

\begin{figure}[t]
\centerline{ \includegraphics[width=0.8\linewidth]{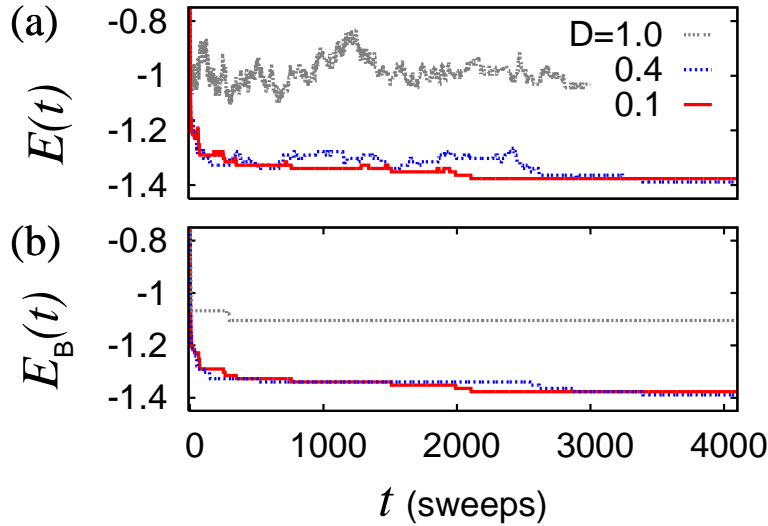}}
\caption{
Sample runs for one test instance of
size $L=18$ and three different values of the diffusion constant
$D$. For $D=0.4$ the GS was correctly identified after $t=3393$
sweeps.  (a) energy of the momentary spin configuration, (b) best
energy value found so far. 
\label{fig:fig5ab}}
\end{figure}  

\begin{figure}[t]
\centerline{ \includegraphics[width=0.8\linewidth]{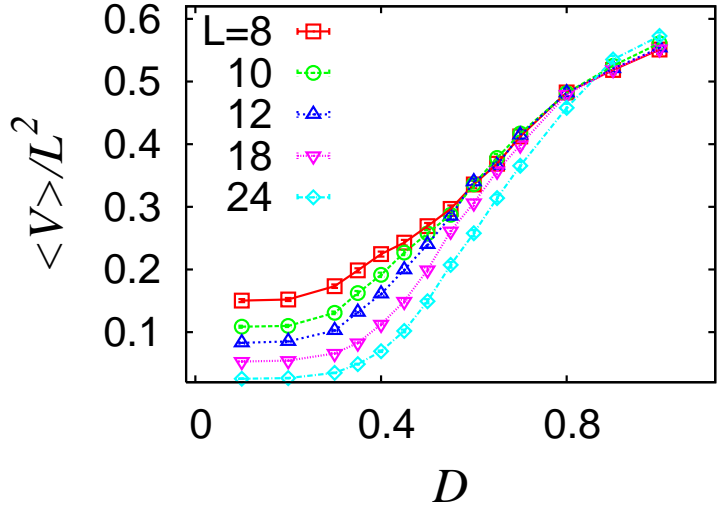}}
\caption{
Cluster analysis of spins that got
flipped at least  once in between two successive values of $E_B(t)$.
Relative size of the largest cluster found for each run.
\label{fig:fig6}}
\end{figure}

To gain insight on why the algorithm performs best near a particular
value of $D$ we analyzed the individual runs more close.  While figure
\ref{fig:fig5ab}(a) shows the momentary configurational energy $E$,
figure \ref{fig:fig5ab}(b) shows $E_B$, the corresponding best energy
value found so far. Disregarding the initial stage of the simulation,
a new lower value for $E_B$ basically means that a different
valley in the energy landscape with a lower value of the minimum
local energy has been found.
 In between two successive records of $E_B$ some
fraction of spins get flipped at least once, which is necessary
to surmount the energy barrier between the valleys. Connected
sets of these spins we call {\em clusters}.
 For each run of the
algorithm we then analyzed the largest of that clusters regarding its
geometric properties, i.e. volume $V$ and spanning lengths with
respect to the independent lattice directions.  So as to limit
boundary effects, we performed the  cluster analysis on test
instances with PP BCs.  Albeit the lattice sizes amenable to numerical
simulation are far too small to  perfom a decent finite-size scaling
analysis we still can use the results to support our intuition on the
performance of the SFP heuristic.
As illustrated in figure \ref{fig:fig6}, simulations reveal that
at small values of $D$ flipped spins are isolated or form clusters
that are negligible compared to the system size. At larger values of
$D$ the flipped spins comprise clusters that cover notable parts of
the whole lattice.
As it appears, a value of the diffusion constant close to $D=0.4$
signifies a threshold above which clusters appear that have a size of
the order of the lattice size.  The occurrence of these clusters is
directly connected to the algorithms capability to surmount energy
barriers in order to escape local minima in the configurational energy.
It is now intriguing to speculate if the performance of the algorithm
is connected to some kind of percolation transition in terms of these
clusters, as observed for  an improved version of extremal
optimization ($\tau$--EO) applied to the $3d$  Edwards-Anderson spin
glass \cite{boettcher2005}.  The data suggests a percolation
transition around $D_c\approx 0.5$ (not shown),  where the value of
$D_c$ was estimated from the common crossing point of the curves  that
describe the percolation probability for different system sizes.
Again, due to the rather small values of $L$ no decent-finite size
scaling analysis  was possible.
For the range of system sizes studied, $D_c$ appears to be somewhat larger
than the optimal value of $D$. Considering the data shown in figure \ref{fig:fig2}
 one finds that the optimal value of $D$ increases slightly as the system 
size increases. Therefore, we cannot rule out that the optimal value of
$D$ shifts to even larger values and approaches $D_c$ in the limit of 
large systems well beyond the maximal system size studied here, i.e. $L=24$.
This would then suggest a connection between the algorithm performance
and a percolation of the investigated clusters.
For the reasons explained above, as this would give an intuitive and plausible
explanation for the optimal performance of the algorithm we consider this 
situation as reasonable.
%
%

\subsection{Dependence on the number of sweeps}
\begin{figure}[t]
\centerline{ \includegraphics[width=1.0\linewidth]{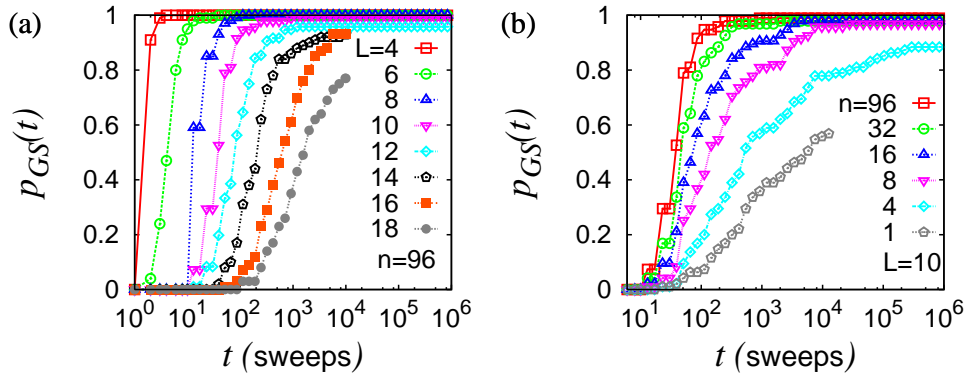}}
\caption{
Fraction of successfully completed test
instances as  function of the number of sweeps carried out at $D=0.4$.  (a)
Increase of typical timescales with increasing  system size $L$, (b)
Decrease of typical timescales with increasing number  $n$ of
simulated copies for each test instance.
\label{fig:fig7ab}}
\end{figure}

We further performed simulations to get a grip on how the fraction of
successfully completed test instances depends on the number of sweeps
carried out by the SFP algorithm. We therein fixed the value of the
diffusion constant to $D=0.4$ so as to ensure an optimal  performance
of the algorithm.  Figure \ref{fig:fig7ab} presents the results
obtained for $100$ test instances for $L=4\ldots18$
each. In the simulation we considered a maximal number of $n=96$
copies for each test instance.
As evident from figure \ref{fig:fig7ab}(a) the typical time scale
after which the SFP algorithm yields a fraction $p_{GS}\approx 1$
increases from $O(1)$ for the smallest system considered,  i.e. $L=4$
($N=16$) to $O(10^3)$ for $L=12$ ($N=144$).  However, an attempt to
characterize quantitatively 
the median running time of the SFP algorithm failed. This
is due to the fact that for several test instances it was not possible
to define a median  running time. Already for $L=6$, $3\%$ of the test
instances did not allow to define a median, meaning that for those
instances less than $n/2$ of the simulated copies resulted in a GS
($12\%$, $18\%$, $34\%$ for $L=8$, $10$, $12$ respectively).
Nevertheless, from this figure is obvious that the effort to
reach a true GS seems to grow exponentially with systems size, although
an exact polynomial algorithm exists. This raises the question,
whether the SFP algorithm can be considered as competitive for
combinatorial optimization problems, see discussion below.
%

Finally,
Figure \ref{fig:fig7ab}(b) illustrates for $L=10$ how the fraction of
correctly identified  GSs is affected by the over all number of copies
simulated for each test instance.  Although the advantage decreases
with increasing $n$ it is beneficial to consider values  $n>1$ since
this significantly reduces the typical time scales on which the SFP
algorithm  approaches $p_{GS}\approx 1$.


\section{Conclusions \label{sect:conclusions}}
We investigated the performance of the recently proposed SFP heuristic, 
applied to the combinatorial optimization problem of finding GSs for
$2d$ $\pm J$ ISGs.
Therefore we compared the results of the heuristic algorithm to those
obtained using exact algorithms, considering systems with up to $N=576$
spin variables. 
Our numerical experiments indicate a best performance of the SFP 
algorithm for a value $D=0.4$ of the diffusion constant. Apart from 
that, the capability of the algorithm to correctly identify GSs rapidly 
decreases with an increasing number of spin variables.
In order to assess its performance we considered a very basic implementation
of the SFP algorithm, meaning that the sampling scheme was carried out at 
a fixed value of the diffusion constant $D$. 
Recently, Rom\'{a} et al. \cite{roma2008} reported on a study that aimed
to determine GS properties of $2d$ and $3d$ ISGs using the parallel tempering
(PT) monte carlo method. For the $\pm J$--model in $2d$ they considered
system sizes up to $L=24$ with PP BCs. For a properly calibrated algorithm 
and considering only $n=1$ independent runs per test instance they obtained
values of $p_{GS}\approx 1.0$ for all the system sizes investigated. This is
superior to the basic implementation of the SFP heuristic studied here, 
where the fraction of correctly identified GSs for $L=12$ already saturates
at a value $p_{GS}\approx0.96$ for $n=96$ ($p_{GS}\approx0.54$ for $n=1$).
Following the paradigm of the
PT algorithm, an improvement of the
method could consist in simulating several copies of one problem realization
at different values of the diffusion constant close to and above the
optimal value $D\approx 0.4$. Copies at distinct values of $D$ should 
therein exchange in a scheduled manner, where the objective is to yield 
spin configurations that possess a minimal configurational energy.
Nevertheless, in contrast to PT, where detailed balance rules,
we do not see in the moment, how such an exchange can be performed
in a controlled manner. Hence, at the current stage, PT approaches
seem to be better heuristics for GSs of combinatorial optimization
problems, in particular also, because PT is quite simple to
implement.

Nevertheless, the SFP is still of interest,
since it is an inherently physical approach and because the optimal performance
may be connected to a percolation transition in the dynamics
of the systems. Hence, the algorithm might contribute to a better
understanding of the relation between dynamic and static complexity
of complex systems.\\
%

\noindent {\emph{Note:}} After submission of our manuscript to {\emph{ J. Stat. Mech.}}
and right before it was accepted for publication, we received 
correspondence from A. Berrones containing valuable comments.
We amended the preprint accordingly, so as to improve the 
manuscript further. To our regret, these amendments do not appear in 
the published version  [{\emph{J. Stat. Mech.}} (2008) P10019] of the Article.

\ack{
We are deeply indebted to L. Saul for providing the computer code to 
obtain the density of states in case of PP BCs.
{We further wish to thank A. Berrones for valuable comments on the manuscript.
We acknowledge financial support from the VolkswagenStiftung (Germany)
within the program  ``Nachwuchsgruppen an Universit\"aten''. The
simulations were performed at the GOLEM I cluster for scientific 
computing at the University of Oldenburg (Germany).
}

\section*{References}

\bibliographystyle{unsrt}
\bibliography{lit_dens_estim.bib}

\end{document}